\documentclass[aps,pra,amsmath,amsfonts,amssymb,twocolumn,nofootinbib]{revtex4}

\usepackage{amssymb}
\usepackage{graphicx,psfrag,color}
\usepackage{dcolumn}
\usepackage{bm}
\usepackage{mathbbol}
\usepackage[T1]{fontenc}

\usepackage{amsthm}
\newtheorem{theorem}{Theorem}

\begin{document}
\flushbottom

\title{Galilean relativity and wave-particle duality imply 
the Schr\"odinger equation}

\author{Gustavo Rigolin}
\email{rigolin@ufscar.br}
\affiliation{Departamento de F\'isica, Universidade Federal de
S\~ao Carlos, 13565-905, S\~ao Carlos, SP, Brazil}

\date{\today}

\begin{abstract}
We show that the Schr\"odinger equation can be derived assuming 
the Galilean covariance of a generic wave equation and the validity
of the de Broglie's wave-particle duality hypothesis. We also obtain from 
this set of assumptions the transformation law for the wave function 
under a Galilean boost and prove that complex wave functions 
are unavoidable for a consistent description of a physical system. 
The extension to the relativistic domain of the above analysis is also
provided. We show that Lorentz covariance and  
wave-particle duality are consistent with two different transformation
laws for the wave function under a Lorentz boost. This leads to two
different wave equations, namely, the Klein-Gordon equation and the
Lorentz covariant Schr\"odinger equation.
\end{abstract}


\maketitle

\section{Introduction} 

Erwin Schr\"odinger arrived at his eponymous equation inspired by
Hamilton's analogy between ordinary mechanics and geometrical optics.
Instead of geometrical optics, 
Schr\"odinger worked with physical (wave) optics 
and searched for its ``mechanical'' analog. This search, guided by 
the de Broglie's wave-particle duality hypothesis \cite{bro25}, 
eventually led to what we know 
today as the Schr\"odinger equation \cite{sch26}.\footnote{We 
should mention that Schr\"odinger was also influenced by 
the second Einstein's paper on the Bose-Einstein condensate
\cite{ein25}, which is acknowledged by Schr\"odinger himself
\cite{sch26b,per10}.}

Our main goal in this work is to derive the Schr\"odinger equation
assuming the de Broglie's wave-particle duality hypothesis and the 
Galilean covariance of the wave equation alone. Furthermore, 
we show that this connection between the Schr\"odinger equation and Galilean covariance is unique, namely, the Schr\"odinger equation is the only linear wave equation satisfying both the de Broglie's hypothesis and 
Galilean covariance.
We also obtain from this set of assumptions the transformation rule 
for the wave function after a Galilean boost
\cite{leb63,leb67,bal98,ang15} and an alternative proof that 
complex wave functions are unavoidable \cite{ren21,li22,che22}. 

The last part of this work deals with the relativistic extension of the
previous analysis. We show that two different classes of 
transformation laws for the wave function are compatible with Lorentz
covariance. One transformation 
law leads to the Klein-Gordon equation \cite{kle26,gor26,foc26,gre00} 
and the other one to the Lorentz covariant Schr\"odinger equation \cite{rig22}. 

Before we move on, we should mention that the Schr\"odinger
equation can be obtained by introducing in
very specific ways stochastic fields or probabilistic arguments into classical physics \cite{nel66,hal02,gri22,bri07,fie11,sch13},
and most of the time by invoking the classical
Hamilton-Jacobi equation \cite{hal02,bri07,fie11,sch13}.
Of particular notice is Ref. \cite{bob95}, where the Schr\"odinger
equation is obtained by assuming 
the existence of a complex wave function that satisfies 
an arbitrary linear wave-equation with at most a first order 
time derivative.

\section{The assumptions}

\subsection{The de Broglie's hypothesis}

The wave-particle duality postulated by de Broglie \cite{bro25} dictates
that any massive particle also has a wave-like character, quantitatively
expressed by the following relations, $\lambda=h/p$ and $\nu=E/h$, 
where $\lambda$ is the particle's ``wavelength'', $\nu$ is  
its ``frequency'', $h$ is Planck's constant, $p$ is the magnitude 
of the particle's momentum $\mathbf{p}$, and $E$ its energy.

In a modern notation, the de Broglie's hypothesis means that a particle's 
energy $E$ and momentum $\mathbf{p}$ are given by
\begin{eqnarray}
E &=& \hbar \omega, \label{E} \\
\mathbf{p} &=& \hbar \mathbf{k} \label{p}, 
\end{eqnarray}
where $\hbar=h/(2\pi)$, the angular frequency $\omega=2\pi\nu$, 
the wave number $k=|\mathbf{k}|=2\pi/\lambda$, and $\mathbf{k}$ is
the wave vector associated with the particle's wave function. 

Throughout de Broglie's PhD thesis \cite{bro25} it is implicit that 
one should look for a wave equation governing the particle's dynamics. This is what Schr\"odinger accomplished three years after 
de Broglie presented 
his PhD thesis \cite{sch26}.  One hundred years later, our main goal
here is to use the de Broglie's wave-particle duality, Eqs. (\ref{E})
and (\ref{p}), plus the following assumption to derive  
the Schr\"odinger equation.

\subsection{Galilean covariance}

In this work, 
Galilean covariance refers to mathematical expressions or physical laws that does not change under 
spatial rotations and Galilean boosts. We will not be dealing with 
space or time translations. These latter two operations
plus spatial rotations and Galilean boosts constitute the inhomogeneous Galilean group. If we exclude space and time translations, we have
the homogeneous Galilean group. Note that Galilean boosts are also known
as Galilean transformations.

If $S$ and $S'$ are two inertial reference frames, with $S'$ moving 
away from $S$ with velocity $\mathbf{v}$, the Cartesian coordinates 
locating a given event and the time of its occurrence in both frames 
are related by the following rules according to the 
Galilean relativity,  
\begin{eqnarray}
t&=& t', \label{g2} \\
\mathbf{r} &=& \mathbf{r'} + \mathbf{v}t'. \label{g1}
\end{eqnarray}
For simplicity and without losing in generality, the above relations 
assume that the origin of time and space coincide in both inertial frames.
We also have, in an obvious notation, that 
$\mathbf{r}=(x,y,z)$ and $\mathbf{r'}=(x',y',z')$ are the space coordinates of the event in $S$ and $S'$, respectively, and 
$t$ and $t'$ are the corresponding time of occurrence of the event.
Equations (\ref{g2}) and (\ref{g1}) are what we call a Galilean boost
or transformation.

Using Einstein summation notation, 
the most general way of writing a wave equation is
\begin{equation}
a^{\mu\nu}\partial_\mu\partial_\nu\Psi(\mathbf{r},t) + b^{\mu}\partial_\mu\Psi(\mathbf{r},t) + f(\mathbf{r},t)\Psi(\mathbf{r},t)=0.
\label{GeneralEq}
\end{equation}
Note that by wave equation we mean the most general
homogeneous linear partial differential equation in the variables 
$x,y,z$, and $t$ of order less or equal to two, and
with constant coefficients multiplying the derivatives. We also assume
that a wave equation has to provide solutions that propagate the waves
along the spacetime, specially for free particles with non zero momentum. 

In Eq.~(\ref{GeneralEq}), $a^{\mu\nu}$
and $b^{\mu}$ are constants (Galilean invariants), 
and $f(\mathbf{r},t)$ is an arbitrary function but a strict Galilean
scalar, namely, $f(\mathbf{r},t)=f'(\mathbf{r'},t')$ after a spatial rotation or a Galilean boost. Also, 
$\partial_\mu=\frac{\partial}{\partial x^\mu}$ and $\mu=0,1,2,3$,
where 
$x^0=ct$, $x^1=x$, $x^2=y$, and $x^3=z$, with $c$ being the speed of 
light in vacuum.

With the previous notation, we can state the principle of Galilean
covariance as follows. If inertial frame $S'$ is connected to 
$S$ by a spatial rotation or a Galilean boost, then if in $S$ the 
wave equation is given by Eq.~(\ref{GeneralEq}), in $S'$ it should look, 
up to an overall non null multiplicative factor,
\begin{equation}
a^{\mu\nu}\partial_{\mu'}\partial_{\nu'}\Psi'(\mathbf{r'},t') 
+ b^{\mu}\partial_{\mu'}\Psi'(\mathbf{r'},t') 
+ f'(\mathbf{r'},t')\Psi'(\mathbf{r'},t')=0.
\label{GeneralEq2}
\end{equation}

\section{Obtaining the Schr\"odinger equation}
\label{III}

\subsection{Covariance under spatial rotations}
\label{rotation}

According to the de Broglie's hypothesis, a particle with mass $m$ has an 
associated wave function describing its dynamics. In the non-relativistic
domain and for a free particle with non null momentum, this wave function
must describe the fact that this particle moves away with 
speed $v=p/m$. Whether or not we deal with a localized particle, or 
with a real or complex wave function, 
variables $\mathbf{r}$ and $t$ are constrained by the following
relation if we want propagating waves,
\begin{equation}
\mathbf{k}\cdot \mathbf{r} - \omega t. \label{propagating}
\end{equation}

Since the scalar product $\mathbf{k}\cdot \mathbf{r}$ is invariant under
spatial rotations, and using that the vacuum is isotropic, Galilean
covariance implies that under any spatial rotation the wave functions 
in frames $S$ and $S'$ are related by the following 
rule,\footnote{If $\mathbf{k}$ and
$\mathbf{r}$ are not related by $\mathbf{k}\cdot \mathbf{r}$, 
rotational covariance can be obtained assuming 
$\Psi(\mathbf{r},t) = g(\mathbf{r'},t) \Psi'(\mathbf{r'},t)$ \cite{rig22}.}
\begin{equation}
\Psi(\mathbf{r},t) = \Psi'(\mathbf{r'},t).
\label{3Drotations}
\end{equation}
Note that any Euclidean vector in $S'$ is connected to its corresponding representation in $S$ by an orthogonal transformation belonging to the $SO(3)$ group, i.e., $\mathbf{r'}=M\mathbf{r}$, with $M \in SO(3)$.
Moreover, if we assume $\Psi(\mathbf{r},t) = \alpha\Psi'(\mathbf{r'},t)$, where $\alpha$ is a constant, it is easy to see that $\alpha=1$. 
Indeed, if we rotate from $S$ to $S'$ and then to $S''$ we get 
$\Psi(\mathbf{r},t) = \alpha^2\Psi''(\mathbf{r''},t)$.
If we rotate directly from $S$ to $S''$ we have
$\Psi(\mathbf{r},t) = \alpha\Psi''(\mathbf{r''},t)$. Comparing both 
expressions for $\Psi(\mathbf{r},t)$ we obtain that $\alpha^2=\alpha$.
Since $\alpha$ cannot be zero, the only valid solution is $\alpha=1$.

Using the transformation law given by Eq.~(\ref{3Drotations}), 
it can be shown that we guarantee covariance under spatial rotations 
if, and only if, for $j,k=1,2,3$ \cite{rig22,rig23a,rig23b},
\begin{eqnarray}
b^j &=& 0, \label{bj}\\ 
a^{j0}+a^{0j} &=& 0,\\
a^{jk}+a^{kj} &=& 0, \hspace{.1cm}\mbox{for}\hspace{.1cm} j\neq k,\\
a^{jj} &=& a^{kk}. \label{ajj}
\end{eqnarray}

Using Eqs.~(\ref{bj})-(\ref{ajj}) and renaming the constant parameters
multiplying the derivatives, Eq.~(\ref{GeneralEq}) becomes
%
%
\begin{equation}
\overline{A}\partial^2_t\Psi(\mathbf{r},t) 
+ \overline{B}\nabla^2\Psi(\mathbf{r},t) 
+ \overline{C}\partial_t\Psi(\mathbf{r},t) 
+ f(\mathbf{r},t)\Psi(\mathbf{r},t)=0.
\label{GeneralEq4}
\end{equation}

Equation (\ref{GeneralEq4}) 
is the most general 
linear partial differential
equation of order two compatible with covariance under
any spatial rotation. To arrive at Eq.~(\ref{GeneralEq4}), 
we used that $\partial_0=\partial^0$, 
$\partial_j=-\partial^j$, for $j=1,2,3$,  
$\partial^2_t=\partial^2/\partial t^2$, and 
$-\partial_j\partial^j=\nabla^2$, where the latter is 
the Laplacian, an invariant under spatial rotations.

\subsection{Covariance under Galilean boosts}
\label{gc}

Since Eq.~(\ref{GeneralEq4}) is covariant under spatial rotations,
we can work without losing in generality with a Galilean boost along
the $x$-axis, where $\mathbf{v}=(v,0,0)$,
\begin{eqnarray}
t&=& t', \label{g2x} \\
x &=& x' + vt', \label{g1x} \\
y &=& y', \label{g3x}\\
z &=& x'. \label{g4x}
\end{eqnarray}

Using the chain rule and Eqs.~(\ref{g2x})-(\ref{g4x}), the first order
derivatives change as follows,
\begin{eqnarray}
\partial_t &=& \partial_{t'} - v \partial_{x'}, \label{delt} \\
\partial_j &=& \partial_{j'}, 
\hspace{.1cm}\mbox{for}\hspace{.1cm} j=x,y,z.\label{delx}
\end{eqnarray}

We also assume that the wave function changes according to the following
rule after a Galilean boost,
\begin{equation}
\Psi(\mathbf{r},t) = g(\mathbf{r'},t')\Psi'(\mathbf{r'},t'),
\label{transformation}
\end{equation}
with $g(\mathbf{r'},t')$ representing an arbitrary function of
$\mathbf{r'}$ and $t'$. Equation (\ref{transformation}) is the most
general way of representing how a wave function changes after an
arbitrary symmetry operation \cite{rig22,rig23a,rig23b}. 

Using Eqs.~(\ref{delt})-(\ref{transformation}), we get that the term multiplying $\overline{A}$ in Eq.~(\ref{GeneralEq4}) is transformed 
to several different terms containing pure or mixed derivatives. 
One of these terms is the following mixed derivative, 
\begin{equation}
-2\overline{A}vg(\mathbf{r'},t')\frac{\partial^2\Psi'(\mathbf{r'},t')}{\partial t'\partial x'}. \label{tx}
\end{equation}

Another term proportional to
$\partial_{t'} \partial_{x'}\Psi'(\mathbf{r'},t')$ cannot
be found anywhere else in the transformed equation after the Galilean boost. The derivatives 
multiplying the other constants in Eq.~(\ref{GeneralEq4}) cannot provide
a mixed derivative in the $t'$ and $x'$ variables if we use 
Eqs.~(\ref{delt})-(\ref{transformation}). Therefore, since $g(\mathbf{r'},t')\neq 0$ and the transformed term given by Eq.~(\ref{tx})
was not in Eq.~(\ref{GeneralEq4}) before the transformation, Galilean
covariance implies that $\overline{A}=0$.
Note that since the boost is along the $x$-axis, $g(\mathbf{r'},t')$
depends only on $x$ and $t$.

This result is interesting by its own since it implies that we cannot have
a covariant equation with a second order derivative in time. In other 
words, we have proved the following result.
%
\begin{theorem}
If the order of the differential equation is at most two, 
Galilean covariance implies that the wave equation cannot have a 
second order time derivative.
\end{theorem}
This is another way of understanding why
the Schr\"o\-dinger equation has only a first order time derivative. 
If higher than second order derivatives are allowed, we show in the Appendix that when fourth order derivatives are present 
we can have a second order time derivative and a covariant 
differential equation. However, for the free particle case, this equation
is essentially the squared Schr\"odinger operator acting on $\Psi$.

Setting $\overline{A}=0$ in Eq.~(\ref{GeneralEq4}), 
Eqs.~(\ref{delt})-(\ref{transformation}) lead to the following 
transformed wave equation if we drop the primes and factor out 
the common term $g(\mathbf{r'},t')$,
\begin{eqnarray}
\overline{B}\nabla^2\Psi + \overline{C}\partial_t\Psi + f\Psi + 
\left[ \frac{2\overline{B}}{g}\frac{\partial g}{\partial x} - \overline{C}v
\right]_1\partial_x\Psi \nonumber \\
+\left[ \frac{\overline{B}}{g}\frac{\partial^2 g}{\partial x^2} 
+ \frac{\overline{C}}{g}\frac{\partial g}{\partial t}
- \frac{\overline{C}v}{g}\frac{\partial g}{\partial x}
\right]_2\Psi=0.
\label{GeneralEq5}
\end{eqnarray}

To obtain covariance, the two brackets above should be zero, i.e.,
$[\hspace{.2cm}]_1=[\hspace{.2cm}]_2=0$. By 
demanding that $[\hspace{.2cm}]_1=0$ and taking the spatial derivative
we get
\begin{equation}
\frac{\overline{B}}{g}\frac{\partial^2 g}{\partial x^2} 
= \frac{\overline{C}v}{2g}\frac{\partial g}{\partial x}. \label{colchete1}
\end{equation}
Using Eq.~(\ref{colchete1}), the second condition for 
covariance, namely, $[\hspace{.2cm}]_2=0$, can be written as
\begin{equation}
\frac{\partial g}{\partial t} = \frac{v}{2}\frac{\partial g}{\partial x}.  
\end{equation}
This is the one-way wave equation whose general solution is
\begin{equation}
g(x,t) = h\left(t+\frac{2}{v}x\right).
\end{equation}
Inserting $g(x,t)$ back into $[\hspace{.2cm}]_1=0$ gives
\begin{equation}
\frac{dh(u)}{du} = \frac{\overline{C}v^2}{4\overline{B}}h(u),
\end{equation}
where $u=t + 2x/v$.  The general solution to the above equation is
\begin{equation}
h(u) = g_0\exp\left[\frac{\overline{C}v^2}{4\overline{B}}u\right],
\end{equation}
with $g_0$ being an arbitrary constant. 
Finally, returning to $g$ and using the definition of $u$ we obtain
\begin{equation}
g(x,t) = g_0\exp\left[\frac{\overline{C}}{4\overline{B}}v^2t
+\frac{\overline{C}}{2\overline{B}}vx\right].
\end{equation}

Putting back the primes, dividing and multiplying the exponent by the 
particle's mass $m$, we arrive at the following solution for a boost in 
an arbitrary direction, 
\begin{equation}
g(\mathbf{r'},t') = g_0\exp\left[\frac{\overline{C}}{2m\overline{B}}
\left(\frac{mv^2}{2}t' + m \mathbf{v}\cdot \mathbf{r'} \right)\right].
\label{grt}
\end{equation}
Note that if we go directly from frame $S$ to $S''$ or from $S$ to $S'$ and then to $S''$, we get that $g_0=1$.

Equation (\ref{grt}) is the most general function $g(\mathbf{r'},t')$
that under a general Galilean boost [cf. Eqs.~(\ref{g2}) and (\ref{g1})] 
guarantees the covariance of the wave equation
(\ref{GeneralEq4}) if its wave function
transforms according to (\ref{transformation}).

If $g(\mathbf{r'},t')$ is a constant, we must 
have $\overline{C}=0$. This implies the following differential 
equation after setting $\overline{A}=\overline{C}=0$ in Eq.~(\ref{GeneralEq4}),
\begin{equation}
\overline{B}\nabla^2\Psi(x) +  f(x)\Psi(x)=0.
\label{EqGconst}
\end{equation}
Note that Eq.~(\ref{EqGconst}) is the 
Helmholtz equation with non constant eigenvalues $f(x)/\overline{B}$. But the most
important point 
is that the above equation has no time 
derivatives. This means that it is not a wave equation at all. Putting
it differently, we have proved the following result. 
%
\begin{theorem}
Galilean covariance implies that the wave function cannot be
a strict scalar under a Galilean boost, i.e., 
it is impossible to have a covariant wave equation 
such that, up to an overall constant phase, 
$\Psi(\mathbf{r},t)=\Psi'(\mathbf{r'},t')$ after a Galilean boost.
\end{theorem}

\subsection{The de Broglie's wave-particle duality}
\label{db}

If we now set $\overline{C}\neq 0$, we have that in general
$g(\mathbf{r'},t')$ is a function of $\mathbf{r'}$ and $t'$.
Moreover, whenever $v\neq 0$, looking at Eq.~(\ref{grt}) we note that 
if $\overline{C}/\overline{B}$ is a positive real number, 
$g(\mathbf{r'},t')$ diverges as a function of the time. And if
$\overline{C}/\overline{B}$ is a negative real number, it tends to 
zero as the time increases. Since  
$\Psi(\mathbf{r},t)=g(\mathbf{r'},t')\Psi'(\mathbf{r'},t')$, this implies
that if $\overline{C}/\overline{B}$ is real and 
$|\Psi'(\mathbf{r'},t')|$ is a non zero bounded function for all $t'$,
we have that 
$|\Psi(\mathbf{r},t)|$ is either zero or infinity 
after a sufficiently long time. 

However, according to the de Broglie's hypothesis, there must be a wave
function with wave number $\mathbf{k}$ and frequency $\omega$ associated
to a particle of mass $m$. For a free particle with non zero velocity, 
the magnitude of this wave function cannot be infinity or zero. Otherwise no physical meaning could be attributed to it and we could not
extract from it a meaningful wave number and frequency that must 
be associated to the moving free particle 
according to the de Broglie's
hypothesis.  
Therefore, we can only satisfy both Galilean covariance and the 
de Broglie's wave-particle duality if $\overline{C}/\overline{B}$ is 
a pure imaginary number. In other words, the wave-particle duality should
be valid in any inertial reference frame and this would not be the case
if we had a zero or divergent wave function. 

The fact that $g(\mathbf{r'},t')$ must be a complex number implies that
we cannot have a real quantum mechanics. Complex numbers, or equivalently
complex wave functions, are mandatory when both Galilean covariance and 
the de Broglie's hypothesis are assumed. This can be proved by the 
following simple argument. Assume that in the inertial reference frame
$S'$, moving away from $S$ with velocity $\mathbf{v}$, we are able to somehow completely describe a particle of mass $m$ using a purely real wave function. If we now go to frame $S$, the particle's wave function
is given by Eq.~(\ref{transformation}). Since $g(\mathbf{r'},t')$ is
necessarily a complex number, in frame $S$ the wave function will thus
be necessarily a complex number too. Therefore, we have proven the 
following interesting result. 
%
\begin{theorem}
Galilean covariance and the de Broglie's wave-particle duality 
imply that complex wave functions are unavoidable in order to properly describe a physical system.
\end{theorem}

To finally arrive at the Schr\"odinger equation, we need to
determine the value of $\overline{C}/\overline{B}$. This is accomplished using
the de Broglie's relations, Eqs.~(\ref{E}) and (\ref{p}), which lead
to a specific dispersion relation for free particles. We then 
require that the wave equation that we have so far, Eq.~(\ref{GeneralEq6}), 
must yield the same 
dispersion relation for its plane wave solution. 
This will fix the value of $\overline{C}/\overline{B}$.

For a non-relativistic free particle of mass $m$ 
moving with constant velocity $\mathbf{v}$, we have that its momentum 
and energy are $\mathbf{p}=m\mathbf{v}$ and  $E=p^2/(2m)$. 
The latter equation together with Eqs. (\ref{E}) and (\ref{p}) 
lead to the following dispersion relation,
\begin{equation}
\omega = \frac{\hbar k^2}{2m}. \label{dispersion}
\end{equation}

The wave equation we have so far can be written as follows,
\begin{equation}
\overline{B}\nabla^2\Psi(\mathbf{r},t) + \overline{C}\partial_t\Psi(\mathbf{r},t) + f(\mathbf{r},t)\Psi(\mathbf{r},t)=0.
\label{GeneralEq6}
\end{equation}
If we insert the ansatz
\begin{equation}
\Psi(\mathbf{r},t) = \Psi_0 e^{i(\mathbf{k}\cdot\mathbf{r}-\omega t)}
\label{ansatz},
\end{equation}
where $\Psi_0$ is an arbitrary constant, into Eq.~(\ref{GeneralEq6})
we get
\begin{equation}
\overline{B} k^2 + i \overline{C} \omega - f(\mathbf{r},t) = 0.
\label{sol1}
\end{equation}
If we now demand that $\omega$ and $k$ are related by Eq.~(\ref{dispersion}), we have that Eq.~(\ref{sol1}) becomes
\begin{equation}
k^2\left( \overline{B} + \frac{i \hbar}{2m}\overline{C} \right) 
- f(\mathbf{r},t) = 0. \label{sol2}
\end{equation}
Since $f(\mathbf{r},t)$ cannot depend on $k$, the only solution to
Eq.~(\ref{sol2}) is
\begin{eqnarray}
\frac{\overline{C}}{\overline{B}} &=& \frac{i2m}{\hbar}, \label{cb} \\
f(\mathbf{r},t) &=& 0. \label{ffree}
\end{eqnarray}

We have fixed the value of $\overline{C}/\overline{B}$ and discovered that for a 
free particle $f(\mathbf{r},t)=0$. With Eq.~(\ref{cb}) we can
write the final expression for the transformation rule for the 
wave equation after a Galilean boost [cf. Eq.~(\ref{grt})],
\begin{equation}
g(\mathbf{r'},t') = \exp\left[\frac{i}{\hbar}
\left(\frac{mv^2}{2}t' + m \mathbf{v}\cdot \mathbf{r'} \right)\right].
\label{grt2}
\end{equation}

Equations (\ref{cb}) and (\ref{ffree})
when inserted into Eq.~(\ref{GeneralEq6}) gi\-ve the Schr\"odinger equation
for a free particle. To write Eq.~(\ref{GeneralEq6}) exactly as we write today the Schr\"odinger  equation, we express $\overline{B}$ and 
$\overline{C}$ as follows, 
\begin{eqnarray}
\overline{B} &=& \frac{\hbar^2}{2m}D, \label{finalB}\\
\overline{C} & = & i\hbar D. \label{finalC}
\end{eqnarray}
Equations (\ref{finalB}) and (\ref{finalC}) satisfy Eq.~(\ref{cb}) and
$D$ is an arbitrary constant. Inserting Eqs.~(\ref{finalB}) and (\ref{finalC}) into (\ref{GeneralEq6}) and using Eq.~(\ref{ffree}),
we obtain the free particle Schr\"odinger equation after dropping the 
overall constant $D$,
\begin{equation}
i\hbar\frac{\partial \Psi}{\partial t} = 
- \frac{\hbar^2}{2m}\nabla^2\Psi.
\end{equation}

We can also fix the value of $f$ in Eq.~(\ref{GeneralEq6}) by studying
the case of a particle in a constant potential $V$ and then generalizing
the obtained equation to a position and time dependent potential. 
For a constant potential, the 
particle's energy is $E=p^2/(2m)+V=\hbar^2k^2/(2m)+V$. The dispersion
relation now is 
\begin{equation}
\omega = \frac{\hbar k^2}{2m} + \frac{V}{\hbar}. \label{dispersionV}
\end{equation}
Inserting Eqs.~(\ref{ansatz}) and (\ref{dispersionV}) into
(\ref{GeneralEq6}), and using Eqs.~(\ref{finalB}) and (\ref{finalC}), 
we get
\begin{equation}
f(\mathbf{r},t)= -D V. \label{finalF}
\end{equation}

Using Eqs.~(\ref{finalB}), (\ref{finalC}), and (\ref{finalF}), we 
can write Eq.~(\ref{GeneralEq6}) after dropping the overall constant
$D$ as follows,
\begin{equation}
i\hbar\frac{\partial \Psi}{\partial t} = 
- \frac{\hbar^2}{2m}\nabla^2\Psi + V\Psi. \label{schrodinger}
\end{equation}
Equation (\ref{schrodinger}) is the celebrated Schr\"odinger equation,
derived here using only two assumptions, namely, Galilean covariance and
the de Broglie's wave-particle duality relations. 

%

\section{Obtaining the relativistic equations}
\label{IV}

Since we will be dealing with relativistic wave equations, it is convenient to use the four-vector notation.
The four-vector notation can be summarized as follows \cite{gre00}. 
The contravariant four-vector $x^\mu$ is defined as 
$(x^0,x^1,x^2,x^3)=(ct,x,y,z)$, where $c$ is the speed of light in 
vacuum. Using the metric where
$g_{00}=1,g_{11}=g_{22}=g_{33}=-1$, with $g_{\mu\nu}=0$ otherwise, 
the covariant four-vector is $x_\mu=g_{\mu\nu}x^\nu$, i.e., 
$(x_0,x_1,x_2,x_3)=(ct,-x,-y,-z)$.
The Einstein summation convention is assumed, where Greek indexes run from $0$ to $3$ while Latin ones go from $1$ to $3$. 
The scalar product between two four-vectors is 
$x^\mu y_\mu$ and  between two spatial vectors is $\mathbf{x}\cdot \mathbf{y}=-x^jy_j$. The covariant four-gradient is defined as 
$
\partial_\mu = 
\left(\frac{\partial}{\partial x^0},\frac{\partial}{\partial x^1},\frac{\partial}{\partial x^2},\frac{\partial}{\partial x^3}
\right)
$
and the contravariant four-gradient by $\partial^\mu=g^{\mu\nu}\partial_\nu$, where $g^{\mu\nu}=g_{\mu\nu}$.

In the four-vector notation, for instance, 
the Galilean boost (\ref{g2}) and (\ref{g1})
can be written as
\begin{eqnarray}
x^0&=& x^{0'}, \label{gl2a} \\
x^j &=& x^{j'} + \beta^j x^{0'}, \label{gl1a}
\end{eqnarray}
where $\beta^j = v^j/c$.
Note that $\beta=|\pmb{\beta}|=|\mathbf{v}|/c=v/c$.

The analysis carried out up to Eq.~(\ref{GeneralEq4}) applies here as
well since we assume for the wave function the same transformation law
under spatial rotations of the non-relativistic case [cf. Eq.~(\ref{3Drotations})]. It is convenient to rewrite Eq.~(\ref{GeneralEq4}) as
\begin{equation}
A\partial^2_0\Psi(x) - B\partial_j\partial^j\Psi(x) + C\partial_0\Psi(x) + f(x)\Psi(x)=0,
\label{GeneralEq3ll}
\end{equation}
where $A=c^2\overline{A}$, $B=\overline{B}$, $C=c\overline{C}$, 
and $x=(x^0,x^1,x^2,x^3)$.

As before, due to the covariance of 
the wave equation (\ref{GeneralEq3ll}) under spatial rotations, we
employ a boost along the $x^1$-axis without losing in generality. However, we must work now with a 
Lorentz instead of a Galilean boost,
\begin{eqnarray}
x^0 &=& \gamma(x^{0'} + \beta x^{1'}), \label{l0} \\
x^1 &=& \gamma(x^{1'} + \beta x^{0'}), \label{l1} \\
x^2 &=& x^{2'}, \label{l2} \\
x^3 &=& x^{3'}, \label{l3}
\end{eqnarray}
where 
\begin{eqnarray}
\beta = \frac{v}{c} & \mbox{and} & \gamma =\frac{1}{\sqrt{1-\beta^2}}. 
\end{eqnarray}

The analogs of Eqs.~(\ref{delt}) and (\ref{delx}) are 
\begin{eqnarray}
\partial_0 &=& \gamma(\partial_{0'} - \beta \partial_{1'}), \label{d0} \\
\partial_1 &=& \gamma(\partial_{1'} - \beta \partial_{0'}), \label{d1} \\
\partial_2 &=& \partial_{2'}, \label{d2} \\
\partial_3 &=& \partial_{3'}. \label{d3}
\end{eqnarray}

Inserting Eqs.~(\ref{transformation}) and (\ref{d0})-(\ref{d3}) 
into Eq.~(\ref{GeneralEq3ll}), and assuming $f(x)$ to be a strict 
Lorentz scalar, $f(x)=f'(x')$, we obtain after factoring out the 
common $g(x')$ term,
\begin{eqnarray}
A'&\hspace{-.1cm}\partial^2_{0'}\Psi'(x') + B'\partial^2_{1'}\Psi'(x') 
+ B \partial^2_{2'}\Psi'(x') + B \partial^2_{3'}\Psi'(x')   \nonumber \\ 
+ &\hspace{-1cm} C'\partial_{0'}\Psi'(x') + D'\partial_{0'}\partial_{1'}\Psi'(x') + E' \partial_{1'}\Psi'(x')
\nonumber \\  
+ &\hspace{-3.0cm}F'\Psi'(x') + f'(x')\Psi'(x') = 0,  
\label{GeneralEq7}
\end{eqnarray}
where
\begin{eqnarray}
A' &=&  \gamma^2(A+\beta^2 B), \label{al} \\
B' &=&  \gamma^2(\beta^2 A + B), \label{bl} \\
C' &=&  \gamma \left[ C + 
2(A+\beta^2 B)\gamma \frac{\partial_{0'}g(x')}{g(x')}\right.
\nonumber \\
&& \left. - 2(A+B)\beta\gamma\frac{\partial_{1'}g(x')}{g(x')}\right],
 \label{cl} \\
D' &=&  - 2\gamma^2\beta (A + B), \label{dl} \\
E' &=&  \gamma \left[ -C\beta - 
2(A+B)\beta\gamma \frac{\partial_{0'}g(x')}{g(x')}\right.
\nonumber \\
&& \left. + 2(B+\beta^2A)\gamma\frac{\partial_{1'}g(x')}{g(x')}\right], \label{el} \\
F' &=& \gamma \left[ C\frac{\partial_{0'}g(x')}{g(x')} + 
(A+\beta^2 B)\gamma \frac{\partial^2_{0'}g(x')}{g(x')}\right.
\nonumber \\
&& \left. - C\beta\frac{\partial_{1'}g(x')}{g(x')}
- 2(A+B)\beta\gamma\frac{\partial_{0'}\partial_{1'}g(x')}{g(x')}\right] \nonumber \\
&& \left. + (B+\beta^2 A)\gamma \frac{\partial^2_{1'}g(x')}{g(x')}
\right] . \label{fl}
\end{eqnarray}

We first note that in Eq.~(\ref{GeneralEq7}), the mixed derivative
term, $D'\partial_{0'}\partial_{1'}\Psi'(x')$, 
should be zero since it is
not present in the wave equation before the Lorentz boost. Therefore,
we must have $D'=0$ which, according to Eq.~(\ref{dl}), implies
\begin{equation}
A = - B. \label{amenosb}
\end{equation}

If we now use Eq.~(\ref{amenosb}) and the mathematical identity
$(1-\beta^2)\gamma^2=1$, Eq.~(\ref{GeneralEq7}) can be written as
\begin{eqnarray}
B\partial_{\mu'}\partial^{\mu'}\Psi'(x') 
 -C'\partial_{0'}\Psi'(x')- f'(x')\Psi'(x')\nonumber \\ 
 -E'\partial_{1'}\Psi'(x')
 -F'\Psi'(x')  = 0,  
\label{GeneralEq8}
\end{eqnarray}
where
\begin{eqnarray}
C' &=&  C\gamma-2B\frac{\partial_{0'}g(x')}{g(x')},\label{cl2} \\
E' &=& -C\beta\gamma+2B\frac{\partial_{1'}g(x')}{g(x')},  
\label{el2} \\
F' &=&  C\gamma\left(\frac{\partial_{0'}g(x')}{g(x')}
-\beta\frac{\partial_{1'}g(x')}{g(x')}\right) \nonumber \\
&&+B\left(\frac{\partial_{1'}^2g(x')}{g(x')}
-\frac{\partial_{0'}^2g(x')}{g(x')}\right). \label{fl2}
\end{eqnarray}

Comparing Eq.~(\ref{GeneralEq8}) with the original differential equation before the Lorentz boost, i.e., comparing it with the equation below, which is Eq.~(\ref{GeneralEq3ll}) when $A=-B$,
\begin{equation}
B\partial_\mu\partial^\mu\Psi(x) - C\partial_0\Psi(x) - f(x)\Psi(x)=0,
\label{GeneralEq3l}
\end{equation}
we realize that to guarantee covariance three conditions must be met,
\begin{eqnarray}
C'&=&C, \label{ccl}\\
E' &=& 0, \label{el0}\\
F' &=& 0. \label{fl0}
\end{eqnarray}

There are two distinct classes of solutions that satisfy 
Eqs.~(\ref{ccl})-(\ref{fl0}). 

\subsection{The Klein-Gordon equation}
\label{kg}

The first class of solutions assumes that it is possible to obtain
a relativistic wave equation such that the wave function is a 
strict scalar under a Lorentz boost, 
i.e., $\Psi(x)=\Psi'(x')$. This is achieved if 
we impose that $g(x')=1$. Note that the results below 
are also true for any non null constant $g(x')$.

For a constant $g$ we immediately see that $F'=0$ since it only 
depends on derivatives of $g$. Furthermore, Eqs.~(\ref{cl2})
and (\ref{el2}) become, respectively, $C'=C\gamma$ and 
$E'=-C\beta\gamma$. We can only have $C'=C$ and $E'=0$, and thus
satisfy Eqs.~(\ref{ccl}) and (\ref{el0}), if $C=0$.

With $C=0$ the wave equation becomes
\begin{equation}
B\partial_\mu\partial^\mu\Psi(x) - f(x)\Psi(x)=0,
\label{GeneralEq9}
\end{equation}
which is the Klein-Gordon equation if we set  
$f(x)$ $=$ $-Bm^2c^2/\hbar^2$.
We can arrive at the previous value for $f(x)$ by demanding 
the validity of the Einstein energy-momentum relation, namely, $E^2=m^2c^4+p^2c^2$, and by applying the 
de Broglie's hypothesis to obtain from the Einstein energy-momentum relation the corresponding dispersion relation  
that the plane wave solution to Eq.~(\ref{GeneralEq9}) must satisfy.

The present analysis can be summarized in the following 
theorem.
\begin{theorem}
Lorentz covariance is compatible with the wave function being
a strict scalar under a Lorentz boost, i.e.,
$\Psi(x)=\Psi'(x')$ after a Lorentz boost. And, together with 
the de Broglie's hypothesis and Einstein energy-momentum
relation, they lead to the Klein-Gordon equation.
\label{teorema4}
\end{theorem}

It is not difficult to see that the logic that led to the above 
theorem can be easily reverted, leading to the following result.
\begin{theorem}
Lorentz covariance, the de Broglie's hypothesis, and 
the Einstein energy-momen\-tum relation imply the Klein-Gordon equation and that the wave function 
is, up to  an overall constant, a strict scalar under a Lorentz boost, i.e., $\Psi(x)=\Psi'(x')$ after a Lorentz boost.
\label{teorema5}
\end{theorem}

The proof of the above theorem is as follows. Lorentz covariance, i.e.,
covariance under spatial rotations and relativistic boosts, leads to 
the wave equation (\ref{GeneralEq3l}) and the auxiliary conditions 
(\ref{ccl})-(\ref{fl0}) that will allow us to obtain the transformation
law for the wave function under a Lorentz boost. If we now use the 
Einstein energy-momentum relation and the de Broglie's wave-particle
hypothesis, we obtain the following dispersion relation,
\begin{equation}
\hbar^2\omega^2=m^2c^4+c^2\hbar^2k^2. \label{rdisp}
\end{equation}
If we use Eq.~(\ref{rdisp}) and the plane wave 
ansatz (\ref{ansatz}), the wave equation (\ref{GeneralEq3l}) 
becomes 
\begin{equation}
\frac{m^2c^2}{\hbar^2}B - \frac{i\omega}{c}C + f(x) = 0.\label{rdisp2}
\end{equation}
Since $B,C$, and $f$ are independent of $k$ and $\omega$, the only 
solution to Eq.~(\ref{rdisp2}) compatible with this constraint is
\begin{eqnarray}
C &=& 0, \label{c=0}\\
f(x) &=& -\frac{m^2c^2}{\hbar^2}B. \label{fc}
\end{eqnarray}
But if $C=0$, Eq.~(\ref{ccl}) implies that $C'=0$. Looking at the
definition of $C'$, Eq.~(\ref{cl2}), this implies that $g(x')$ should not depend on $x^{0'}$. Similarly, Eqs.~(\ref{c=0}), (\ref{el0}) and (\ref{el2}) imply that $g(x')$ should not depend on $x^{1'}$. In other
words, $g(x')$ is a constant, proving that $\Psi(x)$ is a strict scalar under a Lorentz boost. Note that if $g(x)$ is a constant, the remaining
constraint, Eq.~(\ref{fl0}), is automatically satisfied. Finally, if we
insert $f(x)$ as given in Eq.~(\ref{fc}) into the wave equation 
(\ref{GeneralEq3l}) and use the fact that $C=0$, 
we obtain after dropping the common factor $B$ the
Klein-Gordon equation.

\subsection{The Lorentz covariant Schr\"odinger equation}

The second class of solutions to Eqs.~(\ref{ccl})-(\ref{fl0}) no 
longer assumes a constant $g(x')$. We start by solving Eq.~(\ref{el0}).
Using Eq.~(\ref{el2}), $E'=0$ leads to the following partial 
differential equation,
\begin{equation}
\frac{\partial g(x^{0'},x^{1'})}{\partial x^{1'}} = \frac{C\gamma\beta}{2B}g(x^{0'},x^{1'}), \label{dif1}
\end{equation}
whose general solution is
\begin{equation}
g(x^{0'},x^{1'}) = h(x^{0'})
\exp\left(\frac{C\gamma\beta}{2B} x^{1'}\right). \label{gx0}
\end{equation}

Using Eq.~(\ref{ccl}), we have that Eq.~(\ref{cl2}) gives the following
partial differential equation,
\begin{equation}
\frac{\partial g(x^{0'},x^{1'})}{\partial x^{0'}} = \frac{C(\gamma-1)}{2B}g(x^{0'},x^{1'}). \label{dif2}
\end{equation}
Inserting Eq.~(\ref{gx0}) into (\ref{dif2}) we get
\begin{equation}
\frac{d h(x^{0'})}{d x^{0'}} = \frac{C(\gamma-1)}{2B}h(x^{0'}), 
\end{equation}
whose solution is
\begin{equation}
h(x^{0'}) = g_0\exp\left[\frac{C(\gamma-1)}{2B}x^{0'}\right], \label{hx0}
\end{equation}
where $g_0= 1$ (see  Sec. \ref{gc}).

Therefore, using Eqs.~(\ref{gx0}) and (\ref{hx0}) we obtain 
\begin{equation}
g(x^{0'},x^{1'}) =
\exp\left\{ \frac{C}{2B}\left[(\gamma-1)x^{0'}+\gamma\beta x^{1'}\right]
\right\}. \label{gx}
\end{equation}

The remaining constraint, Eq.~(\ref{fl0}), is automatically satisfied
if we use the solution above, Eq.~(\ref{gx}), and Eq.~(\ref{fl2}). It is
worth mentioning that we recover the case of a constant $g(x')$ if 
$C=0$, which we proved to be the case by a different route in 
Sec. \ref{kg}.

Looking at Eq.~(\ref{gx}), we realize that the same analysis carried out
for the non-relativistic case concerning the complex nature of $g(x')$
applies here if $g(x')$ is not a constant ($C\neq 0$). 
This implies that $C/B$ must be a pure imaginary number if $C\neq 0$
and that complex wave functions are unavoidable [cf. Sec. \ref{db}]. 
Note that if $g(x')$ is a constant, the previous
analysis does not apply. This implies that 
real wave functions are compatible with Lorentz covariance and the 
Klein-Gordon equation.

Without further input, $C/B$ is an arbitrary pure imaginary number. 
We can fix its value by demanding that the non-relativistic limit 
($\beta=v/c\ll 1$) of Eq.~(\ref{gx}) tends to Eq.~(\ref{grt2}),
its non-relativistic analog.   

The non-relativistic limit of Eq.~(\ref{gx}) is obtained by expanding its exponent in powers of $v/c$ and keeping only the dominant terms. Since
$\gamma - 1 \approx v^2/(2c^2)$ and $\gamma \beta \approx v/c$, 
Eq.~(\ref{gx}) becomes 
\begin{equation}
g(x',t') \approx
\exp\left[ \frac{C}{2Bc}\left(\frac{v^2}{2}t'+ v x'\right)\right].
\label{gx2}
\end{equation}
We have used that $x^{0'}=ct'$ and $x^{1'}=x'$ to arrive at
Eq.~(\ref{gx2}). Note that $x$ can be the shorthand notation 
for the four-vector $(x^0,x^1,x^2,x^3)$  or simply
the variable associated with the $x$-axis. 
The context makes it clear
which meaning one should attribute to $x$. 
Comparing Eqs.~(\ref{gx2}) and (\ref{grt2}),
where we set $\mathbf{v}=(v,0,0)$ in the latter equation, we realize that
they are equal if
\begin{equation}
\frac{C}{B} = \frac{i2mc}{\hbar}. \label{cb2}
\end{equation}

Inserting Eq.~(\ref{cb2}) into (\ref{gx}) we get
\begin{equation}
g(x^{0'},x^{1'}) =
\exp\left\{ \frac{i}{\hbar}\left[(\gamma-1)mcx^{0'}+\gamma mv x^{1'}\right]\right\}. \label{gx3}
\end{equation}

If we use that $x^{0'}=ct',x^{1'}=x$, and note that 
$vx = \mathbf{v}\cdot\mathbf{r}$, we obtain from Eq.~(\ref{gx3}) the
transformation law for a Lorentz boost in an arbitrary direction 
(along the direction of $\mathbf{v}$), 
\begin{equation}
g(\mathbf{r'},t') = 
\exp\left\{ \frac{i}{\hbar}\left[(\gamma -1)mc^2t'+ 
\gamma m\mathbf{v}\cdot \mathbf{r'}\right]\right\}.
\label{gx4}
\end{equation}

We are still left with one free parameter to fix the values of $C$ and $B$, provided we respect Eq.~(\ref{cb2}). Employing
the same convention of the non-relativistic case, namely, 
Eqs.~(\ref{finalB}) and (\ref{finalC}), we obtain the following 
wave equation from (\ref{GeneralEq3l}), 
\begin{equation}
\partial_\mu\partial^\mu \Psi 
- i\frac{2mc}{\hbar}\partial_0 \Psi + \frac{2mV}{\hbar^2}\Psi=0.
\label{GeneralE10}
\end{equation}
Note that we have also used the non-relativistic convention, 
Eq.~(\ref{finalF}), to rename $f(x)$. In the present case, $V$
should be interpreted as a relativistic invariant or a 
relativistic scalar under proper Lorentz transformations whose 
non-relativistic limit tends to the Newtonian potential energy \cite{rig22}. 

Equation (\ref{GeneralE10}) is the Lorentz covariant Schr\"odinger 
equation obtained in Ref. \cite{rig22} by a different set of assumptions. It can also be written as follows, akin to the way we write the 
non-relativistic Schr\"odinger equation,
\begin{equation}
-\frac{\hbar^2}{2mc^2}\frac{\partial^2\Psi}{\partial t^2} + i\hbar\frac{\partial \Psi}{\partial t} =  
-\frac{\hbar^2}{2m}\nabla^2 \Psi  + V\Psi.
\label{GeneralEq11}
\end{equation}
Looking at Eq.~(\ref{GeneralEq11}) we realize that it differs from the 
non-relativistic Schr\"odinger equation by its first term. It is also 
clear that when $c\rightarrow \infty$, another way to obtain the 
non-relativistic limit, this term goes to zero and we recover the 
Schr\"odinger equation exactly.

Theorems \ref{teorema4} and \ref{teorema5} indicate that the plane
wave solution to Eq.~(\ref{GeneralE10}) does not satisfy the 
dispersion relation (\ref{rdisp}), unless $m=0$. In this particular
case $g(x')$ becomes a constant according to Eq.~(\ref{gx4}) and 
the assumptions that led to those theorems are fulfilled.

In order to better understand this point, let $S'$ be the rest frame
of a particle with mass $m$ that is moving with constant speed 
$\mathbf{v}$ with respect to an inertial frame $S$. 
For the free particle case ($V=0$), we have that in $S'$ the wave
function $\Psi'(\mathbf{r'},t')=\Psi_0$, with $\Psi_0$ being a constant, 
is a solution to Eq.~(\ref{GeneralE10}) that has a clear physical meaning.
Note that this is not the case for the Klein-Gordon equation, which does not accept a constant solution if the mass is not zero. 

A constant solution in $S'$ means a plane wave with a null wave vector and 
zero frequency. As such, the same interpretation of
the non-relativistic case applies here, where we should understand 
$\hbar\omega$ as the kinetic energy of the particle. This becomes even
clearer if we use the transformation rule (\ref{transformation}) and Eq.~(\ref{gx4}) to obtain the wave function in $S$ 
from the one in $S'$, namely, 
\begin{eqnarray}
\hspace{-.5cm}\Psi(\mathbf{r},t)&=&\Psi_0 \exp\left\{ \frac{i}{\hbar}
\left[(\gamma -1)mc^2t'+ \gamma m\mathbf{v}\cdot \mathbf{r'}\right]\right\}, \nonumber \\
&=& \Psi_0 \exp\left\{ \frac{i}{\hbar}
\left[-(\gamma -1)mc^2t+ \gamma m\mathbf{v}\cdot \mathbf{r}\right]\right\}. \label{plane1}
\end{eqnarray}
To obtain the last line, we used the inverse of the Lorentz boost
given by Eqs.~(\ref{l0})-(\ref{l3}) to express the primed variables
as functions of the unprimed ones.

Comparing Eq.~(\ref{plane1}) with the standard way of writing a plane wave, Eq.~(\ref{ansatz}), we recognize that 
\begin{eqnarray}
\hbar \omega &=& (\gamma -1)mc^2, \\
\hbar \mathbf{k} &=& \gamma m\mathbf{v}.
\end{eqnarray}
Note that $(\gamma -1)mc^2$ is the usual relativistic kinetic energy of 
the particle and $\gamma m\mathbf{v}$ its relativistic momentum
from the point of view of $S$. This is consistent with the 
interpretation we gave above about the constant solution in the 
particle's rest frame, where both the kinetic energy 
and momentum are zero.

A detailed analysis of Eq.~(\ref{GeneralEq11})
in several different scenarios and external potentials is 
given in Ref. \cite{rig22} as well as its connection with the 
Klein-Gordon equation. The second quantization of Eq.~(\ref{GeneralEq11})
is also provided in Ref. \cite{rig22} as well as a generalized 
Lorentz covariant Schr\"odinger equation, in which the transformation
law for the wave function under a boost and under spatial rotations 
are given by a non-constant $g(x')$. The extension of the present 
ideas for a spin-$1/2$ elementary particle is
given in Refs. \cite{rig23a,rig23b}, where the first and second quantized
theories are developed. The common feature of the quantum field theories
built on Eq.~(\ref{GeneralEq11}) \cite{rig22} and on its spin-1/2 
extension \cite{rig23a,rig23b} can be summarized in the fact that  
particles and antiparticles with the same mass 
do not have the same dispersion relation anymore. 
This points to a fully relativistic way of understanding the asymmetry 
between matter and antimatter in the present day universe 
\cite{rig22,rig23a,rig23b}.

\section{Conclusion}
\label{conclusion}

We showed that it is possible to derive the Schr\"odin\-ger equation
using only two assumptions, namely, the de Broglie's wave-particle 
duality hypothesis and the Galilean covariance of the wave equation.
The first assumption means that there is a wave function associated 
with a massive particle and that its energy and momentum are connected 
to the frequency and wave vector of that wave as prescribed by 
de Broglie. The second assumption
is the Galilean relativity principle which, in the sense employed in
this work, postulates that the wave equation (laws of physics) 
should look the same after either a spatial rotation or a Galilean 
transformation (Galilean boost).

The above analysis not only led unambiguously to the non-relativistic
Schr\"odinger equation but also to the transformation law for the wave
function after a Galilean boost. It also led to the proof that 
the wave function cannot be a strict scalar under a Galilean transformation, namely, it is not possible to have a non-relativistic
wave equation satisfying the two assumptions outlined above such
that $\Psi(\mathbf{r},t) \rightarrow \Psi(\mathbf{r},t)$ 
after a Galilean boost. 

Furthermore, we showed that 
Galilean covariance and the de Broglie's wave-particle duality also 
imply that complex wave functions are unavoidable for a consistent 
description of a physical system in all inertial frames. We also
showed that any wave equation compatible with those two assumptions cannot have a second order time derivative, which is an alternative way
of understanding why the Schr\"odinger equation has only 
a first order time derivative \cite{bob95}.

The extension of the above results to the relativistic domain was 
given in the end of this work. We derived the wave equations compatible
with the de Broglie's wave-particle duality hypothesis and 
Lorentz covariance, where the latter means covariance under spatial rotations and Lorentz boosts. 

We showed that Lorentz covariance is compatible with 
a wave function that transforms under a Lorentz boost as 
a strict scalar, i.e., 
$\Psi(\mathbf{r},t) \rightarrow \Psi(\mathbf{r},t)$ after a Lorentz transformation.
Moreover, we showed that Lorentz covariance plus 
the de Broglie's hypothesis and the Einstein energy-momentum 
relation lead unambiguously to the Klein-Gordon equation if, and 
only if, $\Psi(\mathbf{r},t) \rightarrow \Psi(\mathbf{r},t)$ under 
proper Lorentz transformations (spatial rotations and boosts).

However, we showed that Lorentz covariance is also compatible with 
a wave function that is not a strict scalar under a Lorentz boost,
namely, with a wave function that transforms after a Lorentz boost 
as follows, 
$\Psi(\mathbf{r},t) \rightarrow g(\mathbf{r},t)\Psi(\mathbf{r},t)$.
By requiring that the non-relativistic approximation of 
$g(\mathbf{r},t)$ should tend to the Schr\"odinger's wave function 
transformation law, we uniquely determined it and also its associated
wave equation. This wave equation was shown to be the Lorentz 
covariant Schr\"odinger equation, derived in Ref. \cite{rig22} by
a different set of assumptions. 

Also, an interesting extension of the
ideas contained in this work would be to search for 
the most general second order linear wave equation and the respective wave function transformation rule 
by requiring ``Einstein's general covariance'' \cite{nor93} for 
the wave equation. This approach may lead to a wave equation 
that incorporates the gravitational field from the start, 
opening the door to a consistent quantum theory of gravity.

Finally, we would like to call attention to an
open problem that we were not able to solve so far, despite several 
attempts. The solution to this problem might lead to an
even deeper understanding of non-relativistic quantum mechanics,
in particular about the origin and meaning of the measurement postulate and Born rule, i.e., the statistical interpretation of 
the wave function. The problem is the following. 
Using the two assumptions outlined above, we were not able to 
derive the Born rule. We believe it is not possible to arrive at
it from those two assumptions alone. However, we do not know either what third physical assumption one should bring to the table to arrive at it. In other words, what is the extra ingredient, 
the extra basic physical law, that we need to prove the Born rule and completely build non-relativistic quantum mechanics 
without ad hoc postulates that do not have a clear physical 
meaning?

\begin{acknowledgments}
GR thanks the Brazilian agency CNPq (National Council for
Scientific and Technological Development) for funding and
the referees for their insightful suggestions. 
\end{acknowledgments}

\appendix*

\section{Higher order differential equations}

\subsection{Third order differential equation}

The most general linear third order partial differential equation
on the variables $x^\mu$ is
\begin{eqnarray}
a^{\mu\nu\kappa}\partial_\mu\partial_\nu\partial_\kappa\Psi(x)
+ a^{\mu\nu}\partial_\mu\partial_\nu\Psi(x) 
+ b^{\mu}\partial_\mu\Psi(x) \nonumber \\
+ f(x)\Psi(x)=0,
\label{thirdorder}
\end{eqnarray}
where we assume that $f(x)\rightarrow f(x)$ after a spatial
rotation or a Galilean boost. In other words, 
$f(x)$ is a strict scalar under those symmetry operations. 

Since the order of the derivatives can be interchanged at will 
(we are assuming that $\Psi(x)$ is a well-behaved infinitely differentiable function), 
we slightly modify the Einstein summation convention
to simplify the analysis. When two or more indexes are summed, we restrict the sum such that each subsequently index is greater or equal to its predecessor. For instance, 
\begin{equation}
a^{\mu\nu}\partial_\mu\partial_\nu = a^{11}\partial_1\partial_1
+ a^{12}\partial_1\partial_2 +  a^{22}\partial_2\partial_2.
\end{equation}

We now start analyzing the constraints on the coefficients $a^{\mu\nu\kappa}$ coming from demanding the covariance of the 
differential equation (\ref{thirdorder}) under spatial rotations. 
We also assume that the wave function is a strict scalar under 
spatial rotations, namely, $\Psi(x) \rightarrow \Psi(x)$ after an
arbitrary spatial rotation. 

The analysis is carried out most simply by separating it into two 
distinct parts. The first one deals with the case where the first
index is $0$, a temporal derivative. The second case considers the 
scenario where only pure spatial derivatives are present. These
two cases exhaust all possibilities.

When $\mu=0$ we have the following third order term,
\begin{equation}
a^{0\nu\kappa}\partial_0\partial_\nu\partial_\kappa\Psi(x).
\end{equation}
Since the first index is fixed and equal to zero and under any
spatial rotation $\partial_0 \rightarrow \partial_0$, what we have
here is an actual two-index term. Therefore, this term and
the other lower order terms 
in Eq.~(\ref{thirdorder}) can be written as
follows after renaming some of the dummy indexes,
\begin{equation}
\overline{a}^{\mu\nu}\partial_\mu\partial_\nu\Psi(x)
+ b^{\mu}\partial_\mu\Psi(x) + f(x)\Psi(x)=0,
\label{effective2}
\end{equation}
where
\begin{equation}
\overline{a}^{\mu\nu}=a^{0\mu\nu}\partial_0+a^{\mu\nu}. \label{term1} 
\end{equation}

According to the results of the main text (see Sec. \ref{rotation}), by demanding the covariance of Eq.~(\ref{effective2}) under spatial rotations we obtain  
\begin{equation}
\overline{a}^{00}\partial_0^2\Psi(x)
-\overline{a}^{11}\partial_j\partial^j\Psi(x)
+ b^{0}\partial_0\Psi(x) + f(x)\Psi(x)=0,
\label{effective2a}
\end{equation}
where we used that $a^{11}=a^{22}=a^{33}$ and  
$\partial_j=-\partial^j$. Inserting Eq.~(\ref{term1}) into 
(\ref{effective2a}) we get
%
%
\begin{eqnarray}
a^{000}\partial_0^3\Psi(x)
-a^{011}\partial_0\partial_j\partial^j\Psi(x)
+a^{00}\partial_0^2\Psi(x) \nonumber \\
-a^{11}\partial_j\partial^j\Psi(x)
+ b^{0}\partial_0\Psi(x) + f(x)\Psi(x)=0.
\label{effective2b}
\end{eqnarray}

When $\mu\neq 0$ we have the following third order term,
\begin{equation}
a^{ijk}\partial_i\partial_j\partial_k\Psi(x),
\label{term2}
\end{equation}
where no temporal index is present. 

If the three indexes are equal, let us say equal to $j$,
a rotation of $\pi$ radians about any one of the other two 
remaining orthogonal axes leads to 
$\partial_j\rightarrow -\partial_j$. This implies that 
$\partial_j^3\rightarrow -\partial_j^3$ and, thus, that 
$a^{jjj}=0$ for $j=1,2,3$. This is true since after the spatial
rotation the term $a^{jjj}\partial_j^3\Psi(x)$ changes to 
$-a^{jjj}\partial_j^3\Psi(x)$. Since the last term was not in the 
differential equation before the transformation, and there are 
other terms in the equation that do not change sign under this 
symmetry operation, we can only guarantee covariance if $a^{jjj}=0$.

If two indexes are equal, a $\pi$ radian rotation about the axis 
labeled by this index will change the sign of the derivative 
related to the index that is different. In other words, 
we will have $a^{jjk}\partial_j^2\partial_k\Psi(x)\rightarrow 
-a^{jjk}\partial_j^2\partial_k\Psi(x)$ and the same argument we
employed above to prove that  $a^{jjj}=0$ leads to $a^{jjk}=0$,
for $j\neq k$.

Finally, if the three indexes are different, we implement a rotation of $\pi/2$ radians about any one of the three orthogonal axes. 
For instance, rotating $\pi/2$ radians about the $k$-axis ($z$-axis),
we have that $\partial_k \rightarrow \partial_k$, 
$\partial_i \rightarrow \partial_j$ and 
$\partial_j \rightarrow -\partial_i$, where $i$ and $j$ means  
the $x$-axis and and $y$-axis, respectively. Therefore, in general,
we will have $a^{ijk}\partial_i\partial_j\partial_k\Psi(x)
\rightarrow -a^{ijk}\partial_i\partial_j\partial_k\Psi(x)$ and thus
by similar arguments already employed in the two previous cases we
must have $a^{ijk}=0$, if $i\neq j \neq k$.

The three cases above combined imply that $a^{ijk} = 0$ for all possible values of $i,j,k$. Therefore, Eq.~(\ref{effective2b}) is 
the most general third order differential 
equation compatible with covariance under spatial rotations.

We need now to impose covariance under Galilean boosts. In this 
scenario the wave function changes according to the following rule,
\begin{equation}
\Psi(x)\rightarrow g(x)\Psi(x), \label{psi}
\end{equation}
where $g(x)\neq 0$.
Note that the shorthand notation for $\Psi(x^0,x^1,x^2,x^3)$ is
$\Psi(x)$. After a 
Galilean boost we also have that
\begin{equation}
\partial_0 \rightarrow \partial_{0} - \beta \partial_{x}, \label{del} 
\end{equation}
where $\beta=v/c\neq0$, while the spatial derivatives remain unchanged (see Secs. \ref{gc} and \ref{IV}).

Using Eqs.~(\ref{psi}) and (\ref{del}), we have that 
$a^{000}\partial_0^3\Psi(x)$ yield after the Galilean boost the following term (among many others, of course),
\begin{equation}
-3\beta g(x) a^{000}\partial_0^2\partial_x\Psi(x).
\label{a000}
\end{equation}
A term proportional to $\partial_0^2\partial_x\Psi(x)$ 
will not appear anywhere else after transforming Eq.~(\ref{effective2b}). This comes about because 
its other terms are either 
at most of second order in the derivatives or 
first order in the time derivative when the order of the 
derivative is higher than two. Therefore, since (\ref{a000})
was not in Eq.~(\ref{effective2b}) before the Galilean 
transformation, we must have that 
\begin{equation}
a^{000}=0 
\label{a000final}
\end{equation}
to guarantee covariance under a Galilean boost.

Moving to the term  
$-a^{001}\partial_0\partial_j\partial^j\Psi(x)$, we
have that after a Galilean boost one of the new terms is
\begin{equation}
\beta a^{001}g(x)\partial_x\partial_j\partial^j\Psi(x).
\label{a001}
\end{equation}
Since $a^{000}=0$ and the other terms in Eq.~(\ref{effective2b}) 
are at most of second order in the derivatives, the fact that a 
term proportional to $\partial_x\partial_j\partial^j\Psi(x)$ is absent before the Galilean transformation implies that we 
necessarily have
\begin{equation}
a^{001}=0 
\label{a001final}
\end{equation}
to preserve the covariance of Eq.~(\ref{effective2b}) under a Galilean boost.

Combining Eqs.~(\ref{a000final}) and (\ref{a001final}), we have that
Eq.~(\ref{effective2b}) becomes
\begin{equation}
a^{00}\partial_0^2\Psi(x)
-a^{11}\partial_j\partial^j\Psi(x)
+ b^{0}\partial_0\Psi(x) + f(x)\Psi(x)=0.
\label{effective2c}
\end{equation}

In other words, if we restrict the order of the derivatives up to three, there is no third order derivative term compatible
with both covariance under spatial rotations and Galilean boosts.
As we show next, we can only have a compatible third order term 
if we allow the presence of fourth order terms as well.

\subsection{Fourth order differential equation}

The same notation, conventions, and assumptions laid out at the previous subsection are valid here. The most general linear 
fourth order partial differential equation
on the variables $x^\mu$ can be written as
\begin{eqnarray}
a^{\mu\nu\kappa\epsilon}\partial_\mu\partial_\nu\partial_\kappa\partial_\epsilon\Psi(x)
+a^{\mu\nu\kappa}\partial_\mu\partial_\nu\partial_\kappa\Psi(x)
\nonumber \\
+ a^{\mu\nu}\partial_\mu\partial_\nu\Psi(x) 
+ b^{\mu}\partial_\mu\Psi(x) 
+ f(x)\Psi(x)=0.
\label{fourthorder}
\end{eqnarray}

The constraints on the coefficients $a^{\mu\nu\kappa\epsilon}$
due to the covariance of the differential equation (\ref{fourthorder}) under spatial rotations are obtained as 
follows. Similarly to the way we worked out the third order
differential equation, we start with the case where the first
index is $0$. The other case, where $\mu\neq 0$, contains only
spatial derivatives and these two cases exhaust all possibilities.

When $\mu=0$ we have,
\begin{equation}
a^{0\nu\kappa\epsilon}\partial_0\partial_\nu\partial_\kappa
\partial_\epsilon\Psi(x).
\end{equation}
Since under any spatial rotation $\partial_0 \rightarrow \partial_0$, we are actually working with an effective three-index
term. This term and the lower order terms of 
Eq.~(\ref{fourthorder}) can be written as follows to highlight this point, 
\begin{eqnarray}
\overline{a}^{\mu\nu\kappa}\partial_\mu\partial_\nu\partial_\kappa\Psi(x)
+a^{\mu\nu}\partial_\mu\partial_\nu\Psi(x)
+ b^{\mu}\partial_\mu\Psi(x)
\nonumber \\
+ f(x)\Psi(x)=0,
\label{effective3}
\end{eqnarray}
where
\begin{equation}
\overline{a}^{\mu\nu\kappa}=a^{0\mu\nu\kappa}\partial_0+
a^{\mu\nu\kappa}. \label{termA} 
\end{equation}  

Following the analysis of the previous subsection, if we require 
the covariance of Eq.~(\ref{effective3}) under spatial rotations we
get that it becomes
\begin{eqnarray}
\overline{a}^{000}\partial_0^3\Psi(x)
-\overline{a}^{011}\partial_0\partial_j\partial^j\Psi(x)
+a^{00}\partial_0^2\Psi(x) \nonumber \\
-a^{11}\partial_j\partial^j\Psi(x)
+ b^{0}\partial_0\Psi(x) + f(x)\Psi(x)=0.
\label{effective3a}
\end{eqnarray}
Inserting Eq.~(\ref{termA}) into 
(\ref{effective3a}) we get
%
%
\begin{eqnarray}
a^{0000}\partial_0^4\Psi(x)
-a^{0011}\partial_0^2\partial_j\partial^j\Psi(x)
+a^{000}\partial_0^3\Psi(x)
\nonumber \\
-a^{011}\partial_0\partial_j\partial^j\Psi(x)
+a^{00}\partial_0^2\Psi(x) 
-a^{11}\partial_j\partial^j\Psi(x)
\nonumber \\
+ b^{0}\partial_0\Psi(x) + f(x)\Psi(x)=0.
\label{effective3b}
\end{eqnarray}

When $\mu\neq 0$ we have the following fourth order term,
\begin{equation}
a^{ijkl}\partial_i\partial_j\partial_k\partial_l\Psi(x),
\label{termB}
\end{equation}
where no temporal index is present.

If the four indexes are equal to $j$, 
a rotation of $\pi/2$ radians about any other 
orthogonal axes not labeled by $j$ leads to 
$\partial_j^4\leftrightarrow \partial^4_k$ or 
$\partial_j^4\leftrightarrow \partial^4_l$, where $j\neq k\neq l$. 
This implies that 
$a^{1111}=a^{2222}=a^{3333}$ to ensure covariance under these 
particular rotations. For example, a counterclockwise rotation of
$\pi/2$ radians about the $x^3$-axis leads to the following changes, 
$a^{1111}\partial_1^4\Psi(x)\rightarrow a^{1111}\partial_2^4\Psi(x)$
and  
$a^{2222}\partial_2^4\Psi(x)\rightarrow a^{2222}\partial_1^4\Psi(x)$,
which implies that $a^{1111}=a^{2222}$ when covariance is required.
A similar analysis changing the axis of rotation gives 
$a^{1111}=a^{3333}$.

If we have three identical indexes, namely, $jjjk$, with $j\neq k$,
a $\pi$ radian rotation about the $x^k$-axis leads to 
$a^{jjjk}\partial_j^3\partial_k\Psi(x)\rightarrow -a^{jjjk}\partial_j^3\partial_k\Psi(x)$ since 
$\partial_j\rightarrow -\partial_j$ and 
$\partial_k\rightarrow \partial_k$ under this rotation. This implies
that $a^{jjjk} = 0$ to guarantee covariance.

If two indexes are equal and the other two are different from 
the latter and from each other, namely, $jjkl$, with $j\neq k\neq l$,
a counterclockwise $\pi$ radian rotation about the $x^k$-axis 
gives $\partial_k\rightarrow\partial_k$, 
$\partial_j\rightarrow -\partial_j$, and
$\partial_l\rightarrow -\partial_l$. This implies that 
$a^{jjkl}\partial_j^2\partial_k\partial_l\Psi(x)\rightarrow 
-a^{jjkl}\partial_j^2\partial_k\partial_l\Psi(x)$ and thus
we must have $a^{jjkl}=0$ to obtain covariance.

Now, if two indexes are equal and the other two are also equal 
but different from the other pair, namely, $jjkk$, with $j\neq k$,
a counterclockwise $\pi/2$ radian rotation about the $x^j$-axis
gives 
$\partial_j^2\rightarrow\partial_j^2$ and 
$\partial_k^2\leftrightarrow\partial_l^2$, 
where $l\neq j$ and $l\neq k$.
This implies that $a^{jjkk}\partial_j^2\partial_k^2\Psi(x)\rightarrow
a^{jjkk}\partial_j^2\partial_l^2\Psi(x)$ while
$a^{jjll}\partial_j^2\partial_l^2\Psi(x)\rightarrow
a^{jjll}\partial_j^2\partial_k^2\Psi(x)$
and thus
we must have $a^{jjkk}=a^{jjll}$ to have covariance.

Combining the results of the last four paragraphs, we obtain that
Eq.~(\ref{termB}) becomes
\begin{equation}
[a^{1111}(\partial_1^4+\partial_2^4+ \partial_3^4)
+a^{1122}(\partial_1^2\partial_2^2+\partial_1^2\partial_3^2
+ \partial_2^2\partial_3^2)]\Psi(x).
\label{termB2}
\end{equation}

We need one last rotation, namely, a counterclockwise $\pi/4$ radian rotation about the $x^3$-axis. In this case 
$\partial_3\rightarrow \partial_3$,
$\partial_1\rightarrow (\partial_1 + \partial_2)/\sqrt{2}$ and
$\partial_2\rightarrow (\partial_2 - \partial_1)/\sqrt{2}$. Therefore,
transforming Eq.~(\ref{termB2}) we get
\begin{eqnarray}
\hspace{-.5cm}a^{1111}\left(\frac{\partial_1^4+\partial_2^4}{2}
+ 3\partial_1^2\partial_2^2 + \partial_3^4
\right) &\hspace{-.3cm}\Psi(x) \nonumber \\
\hspace{-.5cm}+a^{1122}\left(\frac{\partial_1^4+\partial_2^4}{4}
-\frac{\partial_1^2\partial_2^2}{2} + \partial_1^2\partial_3^2
+ \partial_2^2\partial_3^2
\right)&\hspace{-.3cm}\Psi(x).
\label{termB3}
\end{eqnarray}

Looking at Eq.~(\ref{termB3}) we note that, individually, the 
operators multiplying $a^{1111}$ and $a^{1122}$ are not covariant
under the latter rotation. However, the whole expression can be 
made covariant if we set
\begin{equation}
a^{1122} = 2 a^{1111}.
\label{a1111final}
\end{equation}

Indeed, using Eq.~(\ref{a1111final}) we have that Eq.~(\ref{termB3})
becomes
\begin{equation}
a^{1111}\left(
\partial_1^2+\partial_2^2+\partial_3^2
\right)^2\Psi(x)=
a^{1111}
\partial_j\partial^j\partial_k\partial^k\Psi(x),
\label{termB4}
\end{equation}
which is the same expression we get by also inserting
Eq.~(\ref{a1111final}) into (\ref{termB2}), where the latter 
is the term under investigation before the transformation. 

Moreover, the operator 
$(\partial_1^2+\partial_2^2+\partial_3^2)^2$
is actually the square of the Laplacian, i.e., 
$(\nabla^2)^2\equiv \nabla^2(\nabla^2)$. And
since the Laplacian operator is invariant under any rotation, we
have that Eq.~(\ref{termB4}) is covariant under any rotation as 
well.

Using Eqs.~(\ref{effective3b}) and (\ref{termB4}), and renaming 
the constant coefficients, we can write
the most general linear fourth order partial differential 
equation that is covariant under arbitrary spatial rotations 
as follows,
\begin{eqnarray}
\widetilde{A}\partial_0^4\Psi(x) 
+\widetilde{B}\partial_j\partial^j\partial_k\partial^k\Psi(x)
-\widetilde{C}\partial_0^2\partial_j\partial^j\Psi(x)
\nonumber \\
+\widetilde{a}\partial_0^3\Psi(x)
-\widetilde{b}\partial_0\partial_j\partial^j\Psi(x)
+A\partial_0^2\Psi(x)
\nonumber \\
-B\partial_j\partial^j\Psi(x)
+ C\partial_0\Psi(x) 
+ f(x)\Psi(x)=0.
\label{effective3c}
\end{eqnarray}

We need now to analyze the covariance under a Galilean boost.
Using Eqs.~(\ref{psi}) and (\ref{del}), we see that one of the 
terms coming from $\widetilde{A}\partial_0^4\Psi(x)$ 
after the boost is
\begin{equation}
-4\beta g(x) \widetilde{A} \partial_0^3\partial_x\Psi(x).
\label{Atil}
\end{equation}
A mixed derivative term proportional to 
$\partial_0^3\partial_x\Psi(x)$ does not come from any other 
transformed term of Eq.~(\ref{effective3c}). 
This is true because 
the other terms in Eq.~(\ref{effective3c}) are either 
at most of third order in the derivatives or 
second order in the time derivative when the order of the 
derivative is higher than three. Hence, since this term is 
absent before the boost, we can only guarantee 
covariance under Galilean transformations if  
\begin{equation}
\widetilde{A}=0. 
\label{Atilfinal}
\end{equation}

If we now use Eqs.~(\ref{psi}) and (\ref{del}), one of the terms 
stemming from 
$-\widetilde{C}\partial_0^2\partial_j\partial^j\Psi(x)$ after the 
boost is 
\begin{equation}
2\beta g(x)\widetilde{C}
\partial_0\partial_x\partial_j\partial^j\Psi(x).
\label{Ctil}
\end{equation}
Since we must have $\widetilde{A}=0$ and the other terms in Eq.~(\ref{effective3c}) 
are either at most of third order in the derivatives or do not 
have a time derivative, a similar term
proportional to $\partial_0\partial_x\partial_j\partial^j\Psi(x)$
will not appear after the boost. Thus, since a term like this is not
present before the transformation, we have
\begin{equation}
\widetilde{C}=0 
\label{Ctilfinal}
\end{equation}
to preserve the covariance of Eq.~(\ref{effective3c}) under a Galilean boost.

A similar argument used to prove Eq.~(\ref{a000final}) applies here 
as well. This leads to  
\begin{equation}
\widetilde{a} = 0 \label{atilfinal}
\end{equation}
in order to preserve covariance under boosts.

Combing the results given by Eqs.~(\ref{Atilfinal}),
(\ref{Ctilfinal}), and (\ref{atilfinal}), the 
differential equation (\ref{effective3c}) becomes
\begin{eqnarray}
\widetilde{B}\partial_j\partial^j\partial_k\partial^k\Psi(x)
-\widetilde{b}\partial_0\partial_j\partial^j\Psi(x)
+A\partial_0^2\Psi(x)
\nonumber \\
-B\partial_j\partial^j\Psi(x)
+ C\partial_0\Psi(x) 
+ f(x)\Psi(x)=0.
\label{effective3d}
\end{eqnarray}
It is worth mentioning that after the boost,
the fourth order term
$\widetilde{B}\partial_j\partial^j\partial_k\partial^k\Psi(x)$
will provide similar terms to those coming from 
$-\widetilde{b}\partial_0\partial_j\partial^j\Psi(x)$. 
This is why the argument used in the previous subsection to
rule out the latter term is no longer valid. Similarly,
since the latter term is now present, after the boost it 
will give similar terms to those coming from 
$A\partial_0^2\Psi(x)$. This is why now the presence of a pure 
second order time derivative will not contradict Galilean covariance
and why the proof of the main text ruling it out 
no longer applies.

Applying Eqs.~(\ref{psi}) and (\ref{del}) to Eq.~(\ref{effective3d}),
after a long but straightforward calculation, similar to 
the ones we explicitly showed in Secs. \ref{III} and \ref{IV}, we get 
that we can only guarantee covariance if
\begin{eqnarray}
\hspace{-1cm}\widetilde{B} &=& \frac{\overline{A}B^2}
{\overline{C}^2},\label{Btilde} \\
\hspace{-1cm}\overline{b} &=& \frac{2\overline{A}B}{\overline{C}}, 
\label{bbar}\\
\hspace{-1cm}
g(\mathbf{r'},t')&=&\exp\left[ \frac{\overline{C}}{2mB}
\left(\frac{mv^2t'}{2}
+m\mathbf{v}\cdot \mathbf{r'}\right)\right],
\label{gnew}
\end{eqnarray}
where $A=c^2\overline{A}$, $C=c\overline{C}$, and 
$\widetilde{b}=c\overline{b}$. We also used that $x^{0'}=ct'$ 
to arrive at the last equation.

To fix the value of $\overline{C}/B$ we require
that the plane wave solution to Eq.~(\ref{effective3d}) 
satisfies the free particle dispersion relation (\ref{dispersion}), where the latter is a consequence of the de Broglie's wave-particle duality hypothesis. Inserting the ansatz
(\ref{ansatz}) into Eq.~(\ref{effective3d}), we obtain the 
dispersion relation (\ref{dispersion}) if
\begin{eqnarray}
\frac{\overline{C}}{B} &=& \frac{i2m}{\hbar}, \label{cb3} \\
f(\mathbf{r},t) &=& 0. \label{ffree2}
\end{eqnarray}
Note that with this value for $\overline{C}/B$, 
$g(\mathbf{r'},t')$ given by Eq.~(\ref{gnew}) is exactly the one
we obtained for the standard Schr\"odinger equation 
[cf. Eq.~(\ref{grt2})]. In other words, the wave functions of 
the fourth order equation and the Schr\"odinger equation have 
the same transformation law under a Galilean boost.

Using Eqs.~(\ref{Btilde}), (\ref{bbar}), and  (\ref{cb3}),
the wave equation can be written as 
\begin{eqnarray}
\frac{\overline{A}\hbar^2}{4m^2}\nabla^4\Psi
+\frac{i\hbar \overline{A}}{m}\nabla^2\partial_t\Psi
-\overline{A}\partial_t^2\Psi
\nonumber \\
-B\nabla^2\Psi
 -\frac{i2mB}{\hbar}\partial_t\Psi 
- f\Psi=0,
\label{effective4}
\end{eqnarray}
where $\nabla^4\equiv \nabla^2(\nabla^2)$, i.e., the square of the 
Laplacian.

It is interesting to note that if $\overline{A}=0$, or equivalently
$A=0$, we recover the standard Schr\"odinger equation by 
properly adjusting the free parameter $B$, i.e., by choosing 
$B=-\hbar^2/(2m)$ and by setting $f=V(\mathbf{r},t)$. Since 
the constant $A$ multiplies the second order time derivative, 
its suppression 
leads immediately to the Schr\"odinger equation, even if we go 
up to fourth order.

Moreover, writing the Schr\"odinger equation as 
\begin{equation}
\mathcal{\hat{S}}\Psi = 0, \label{schrodinger2}
\end{equation}
where
\begin{equation}
\mathcal{\hat{S}} = 
\frac{\hbar^2}{2m}\nabla^2 
+ i\hbar\partial_t - V
\end{equation}
is what we call the Schr\"odinger operator, we obtain for
a constant $V$ that
\begin{equation}
\mathcal{\hat{S}}^2 = \frac{\hbar^4}{4m^2}\nabla^4 
+\frac{i\hbar^3}{m}\nabla^2\partial_t - \hbar^2\partial^2_t
-V\frac{\hbar^2}{m}\nabla^2 - i2V\hbar\partial_t
+V^2. \label{s2}
\end{equation}

Comparing Eq.~(\ref{s2}) with the operator acting on $\Psi$ that 
generates the wave equation (\ref{effective4}), we realize that
they are equal if
\begin{eqnarray}
\overline{A} &=& \hbar^2, \label{Abar}\\
B & = & \frac{V\hbar^2}{m}, \\
f & = & -V^2. \label{f}
\end{eqnarray}

In other words, for a free particle ($V=0$), or a particle in a
constant potential, 
the fourth order wave equation (\ref{effective4}) is 
obtained from ``squaring'' the Schr\"odinger equation. 
Specifically, it is obtained
by squaring the Schr\"odinger operator $\mathcal{\hat{S}}$.
The Schr\"odinger equation is given by $\mathcal{\hat{S}}\Psi=0$
and the fourth order equation by $\mathcal{\hat{S}}^2\Psi=0$ if
$\overline{A}$, $B$, and $f$ are
given by Eqs.~(\ref{Abar})-(\ref{f}). The latter relation also
implies that any solution to the Schr\"odinger equation is 
also a solution to the fourth order equation. Indeed, if $\Psi$
is a solution to the Schr\"odinger equation we have 
$\mathcal{\hat{S}}\Psi=0$. But  
$\mathcal{\hat{S}}^2\Psi=\mathcal{\hat{S}}(\mathcal{\hat{S}}\Psi)$.
Therefore, we must have $\mathcal{\hat{S}}^2\Psi=0$, proving that
$\Psi$ is also a solution to the fourth order equation.

Finally, if $V$ is not constant but depends on the position, namely,
if $V=V(\mathbf{r})$, one of the terms of the square of the Schr\"odinger operator is
\begin{equation}
-\frac{\hbar^2}{m}\nabla V\cdot \nabla\Psi.
\end{equation}

Looking at Eq.~(\ref{effective4}), we realize that this type of 
mixed derivative is absent from it. Therefore, it is not possible,
in general, to make the square of the Schr\"odinger equation 
($\mathcal{\hat{S}}^2\Psi=0$) equivalent to the fourth order 
wave equation (\ref{effective4}) if $V$ is not a constant.

\end{document}